\documentclass[twocolumn]{aastex62}
\usepackage{amsmath}

\renewcommand{\sc}{{\rm sc}}

\renewcommand{\vec}[1]{\mathbf{#1}}

\newcommand{\aave}[1]{\left\langle #1\right\rangle}

\newcommand{\rms}{{\rm r.m.s.}}
\graphicspath{{./}{figures/}}

\shorttitle{Interpreting solar wind turbulent spectra beyond TH}
\shortauthors{Bourouaine \& Perez}

\begin{document}

\title{Interpreting solar wind turbulent spectra beyond Taylor's Hypothesis}

\correspondingauthor{Sofiane Bourouaine}
\email{sbourouaine@fit.edu}

\author[0000-0002-0786-7307]{Sofiane Bourouaine}
\affil{Florida Institute of Technology \\
150 w Univeristy Blvd  \\
Melbourne, Fl, 32904, USA}

\author{Jean C. Perez}
\affil{Florida Institute of Technology \\
150 w Univeristy Blvd  \\
Melbourne, Fl, 32904, USA}

\begin{abstract}

In this paper we apply a methodology, recently proposed by Bourouaine and Perez (2019) (BP19), to interpret solar-wind turbulent power spectra beyond Taylor approximation (TA). The turbulent power spectra were measured using \emph{Helios} spacecraft data near 0.6 au. We use the model proposed in BP19 to reproduce the field-perpendicular power spectrum $E(k_\perp)$ of anti-sunward Alfv\'enic fluctuations in the plasma frame (where $k_\perp$ is the field-perpendicular wavenumber) from the corresponding measured frequency power spectrum $P_{\sc}(\omega,\theta_b)$ along the sampling angle $\theta_b$, which is the angle between the local magnetic field and the sampling direction. Here $\omega=2\pi f$  and $f$ is the frequency of the time signal. Interestingly enough, we found that for all corresponding measured frequency power spectrum $P_{\sc}(\omega,\theta_b)$ the reproduced field-perpendicular power spectrum $E(k_\perp)$ is the same and independent of the considered sampling angle $\theta_b$. This finding is consistent with the fact that the analyzed turbulence is strong and highly anisotropic with $k_\| \ll k_\perp$ (where $k_\|$ is the field-parallel wavenumber). Furthermore, for this specific time signal we found that the commonly used TA is still approximately valid with the important difference that a broadening in $k_\perp$ for each angular frequency $\omega$ is present. This broadening can be described in the context of the methodology proposed in BP19.

\end{abstract}

\keywords{}

\section{Introduction} \label{sec:intro}

Most analyses of solar wind observations normally adopt the Taylor approximation (TA), often called Taylor's hypothesis (TH)~\citep{taylor38}, to investigate turbulent time signals from in-situ spacecraft measurements, and thus study the fundamental physics of solar wind turbulence. TH assumes that when the turbulent fluctuations are advected with a speed $U$ that is much higher than the typical fluctuation speed $v$ ($U\gg v$), the time ($\tau$) and the spatial ($\bf r$) lags of the measured structures are connected as ${\bf r}={\bf U}\tau$. However, this frozen-in-flow approximation may not be applicable in all solar wind conditions, most importantly at heliocentric distances where the recently launched Parker Solar Probe (\emph{PSP})~\citep{fox16} mission is expected to explore, see for instance~\cite[][BP18 hereafter]{bourouaine18}.

Recently, with the new \emph{PSP} mission there has been an increased and renewed interest in investigating the applicability of the TH in the solar wind~\citep{klein14,klein15,narita17,bourouaine18,bourouaine19,huang19,chhiber19}. More recently, Bourouaine and Perez (2019) (BP19) proposed a phenomenological model to explain the time decorrelation of the turbulent structures by extending the hydrodynamic Kraichnan's sweeping model to MHD turbulence. The form of the temporal part of the two-point two-time correlation function in the BP19 model allows for the interpretation of the turbulent time signal even when $v\sim U$ and TH is not valid.
 
The fundamental physics of the time decorrelation of the MHD turbulent structures has been investigated in a number of previous works~\citep{matthaeus10,matthaeus16,servidio11,narita13,narita17,weygand13}. For instance, BP19 found that the Eulerian decorrelation in simulations is consistent with spectral broadening associated with pure hydrodynamic sweeping by the large-scale eddies, combined with a Doppler shift associated with Alfv\'enic propagation along the background magnetic field in strong MHD turbulence. 

In this letter we present a practical application of the model proposed in BP19 to reproduce the energy spectrum in the plasma frame beyond the TA. For this purpose we analyze turbulent signals measured by \emph{Helios 2} near 0.6 au. In section 2 we calculate the power spectra of the anti-sunward propagating Elssaser field and the fluid velocity field in the spacecraft frame through the correlations functions. In section 3, we reproduce the reduced field-perpendicular power spectrum and the  broadening in the field-perpendicular wavenumber $k_\perp$, using the model proposed in BP19 , and then using the TA. Finally, we summarize and discuss the obtained results in section 4.

\section{Data and methodology} \label{sec:cite}

\begin{figure}[!t]
    \centering
    \includegraphics[width=0.5\textwidth]{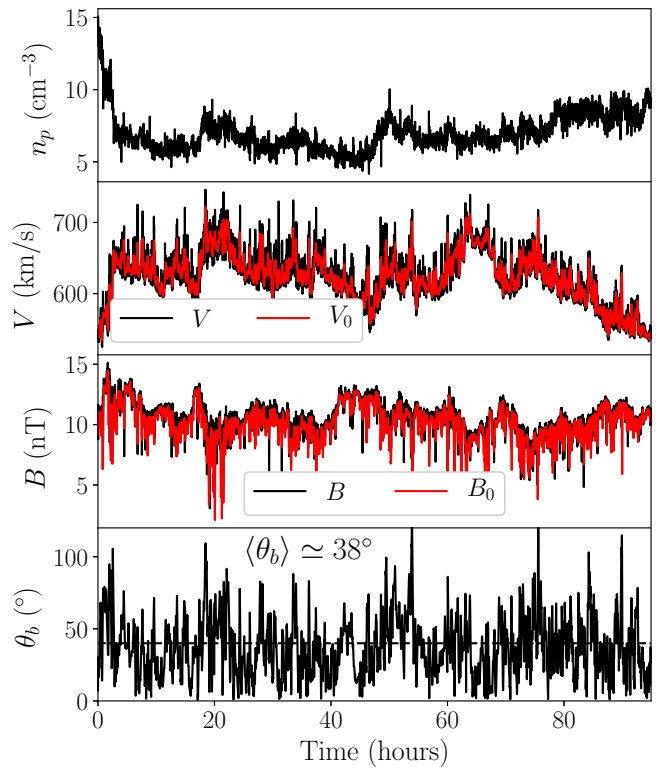}
     \vspace*{-0.1cm}
    \caption{Time period of the signal is from date 03/15/1976 (00:00:30.00) to 03/18/1976 (22:58:12.00) near 0.6 au.
    Upper panel: proton number density. Upper middle panel: solar wind speed $V$ (black line) with local mean speed $V_0$ (red line). Lower middle panel: magnetic field magnitude $B$ (black line) with local mean field $B_0$ (red line). Lower panel: sampling angle $\theta_b$ between vectors $\vec V_0$ and $\vec B_0$. All signals are plotted as functions of time $t$.}
    
    \label{fig:fig1}
\end{figure}

In our analysis we use combined plasma-field data provided by \emph{Helios 2} with a time resolution of about 40.5~s. We focus on the analysis of plasma and magnetic field signals measured within the  time period 03/15/1976 (00:00:30.00) to 03/18/1976 (22:58:12.00). During this time period the spacecraft is passing mostly through fast solar wind as we can see in Figure~\ref{fig:fig1}.


In this analysis we aim to reproduce the reduced energy spectrum $E(k_\perp)=2\pi k_\perp\int h_0(k_\parallel,k_\perp) dk_\parallel$  in the inertial range. Here $h_0(k_\parallel,k_\perp)$ is the plasma-frame three-dimensional power spectra corresponding to the anti-sunward propagating Elsasser field $\delta\vec z^{-}=\delta\vec v- \delta\vec b/\sqrt{4\pi\rho}$, where $\delta\vec v$ is the fluctuating fluid velocity vector, $\delta\vec b$ is the fluctuating magnetic field vector and $\rho$ is the proton mass density. We define the local mean magnetic and velocity vectors through the moving average over a period $T$ around time $t_j$, i.e.,
\begin{eqnarray}
\vec B_0(t_j)&=& \frac 1{N_j}\sum_{i}W_T(t_j-t_i)\vec b(t_i)\\
\vec V_0(t_j)&=& \frac 1{N_j}\sum_{i}W_T(t_j-t_i) \vec v(t_i)
\end{eqnarray}
where $N_j$ is the number of averaging samples and $W_T(t)$ is a windowing function that vanishes everywhere except at $|t|\le T/2$, in which case it is equal to one. The period $T\simeq 9$ min is chosen to be close to largest scale within the inertial range (Figure~\ref{fig:fig3}). The magnitudes of the local mean velocity and local magnetic field are shown in red color lines in Figure~\ref{fig:fig1}. The corresponding angle $\theta_{b}(t)$ between the two vectors, $\vec V_0(t)$ and $\vec B_0(t)$ is plotted as a function of time $t$ in the bottom panel of Figure~\ref{fig:fig1}. 

In the following analysis we estimate the power spectra $P_{\sc}(\omega, \theta_i)$ and $P^{v}_{\sc}(\omega, \theta_i)$ that correspond to $\vec z^{-}$ and $\vec v $, respectively, along a sampling angle $\theta_b\simeq \theta_i$, where $\theta_i$ are the angle bins of width $\Delta\theta=10^\circ$ centered at the following angle values $\theta_{0}=20^{\circ}$, $\theta_{1}=30^{\circ}$, $\theta_{2}=40^{\circ}$, $\theta_{3}=50^{\circ}$ and $\theta_{4}=60^{\circ}$.


Generally, the power spectrum of a fluctuating  vector quantity $\vec a(t)$ can be obtained through the Fourier transform of its auto-correlation function $R(\tau)=\aave{ \vec a(t_0)\cdot \vec a(t_0+\tau)}$ (see e.g., \cite{bourouaine13}), where $\tau$ is the time-lag. In our analysis, the empirical power spectra $P_{\sc}(\omega, \theta_i)$ and $P^{v}_{\sc}(\omega, \theta_i)$ are then obtained as the Fourier transform of the following conditioned correlations functions
\begin{eqnarray}
C_v(\tau,\theta_i)&=&\aave{\left(\vec v(t)-\overline{\vec v}\right)\cdot \left(\vec v(t+\tau)-\overline{\vec v}\right)}_{\theta_i,V_0, R}, 
\label{Eq.Corrvper}\\
C_z(\tau,\theta_i)&=&\aave{\left(\vec z^{-}(t)-\overline{\vec z}^{-}\right)\cdot \left(\vec  z^{-}(t+\tau)-\overline{\vec z}^{-}\right)}_{\theta_i,V_0,R}
\label{Eq.Corrb}
\end{eqnarray}
where $\aave{\cdots}_{\theta_i,V_0,R}$ denotes the ensemble average, which can be computed over many realizations (or average over time $t$) conditioned by the angle bin $\theta_i$, mean velocity $V_0$ and the transverse ratio $R=(\delta z^{-}_\parallel/\delta z^{-}_\perp )^2$. Here, the perpendicular and the parallel components of the fluctuations $\delta \vec z^{-}$ 
are defined with respect to the local mean magnetic field $\vec B_0$. As we are interested in fast solar wind and transverse fluctuations, we calculate correlation functions by considering only the statistics of those two times $t$ and $t+\tau$ in Equations~\eqref{Eq.Corrvper}~and~\eqref{Eq.Corrb} for which the corresponding values of mean velocity $V_0\ge 600$ km/s and the ratio $R < 0.2$.  $\overline{\vec v}$ and $\overline{\vec z}^{-}$ were obtained through averaging over all considered points in the calculation of the correlation functions. 

Note that the correlation functions in Eqs.~\eqref{Eq.Corrvper}~and~\eqref{Eq.Corrb} are calculated using global mean vectors $\overline{\vec v}$ and $\overline{\vec z}^{-}$ instead of local mean vectors $\vec V_0$ and $\vec B_0$, respectively. This is done in order to capture the frequency power spectrum of the outer scale (i.e., scales that are larger than period $T$), which then allows us to properly estimate the root mean squared (\rms)~speed of the energy-containing eddies required for the reproduction of the reduced energy spectrum $E(k_\perp)$ according to BP19 model.

It is worth mentioning that there are two main advantages of calculating the power spectra through the correlation functions, 1) we can be selective and avoid any unwanted points, including gaps of bad measurements in the calculation of the correlation functions, and
2) we can check the statistics that correspond to the estimation of the correlation functions for each time-lag $\tau$ including the statistics of the outer scale for large $\tau$.

\begin{figure}
    \centering
    \includegraphics[width=0.5\textwidth]{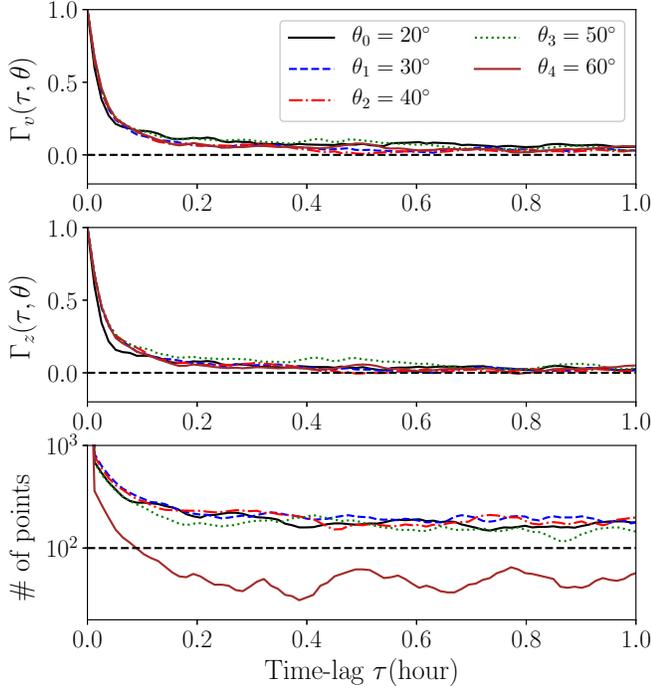}
       \vspace*{-0.5cm}
    \caption{Normalized correlation functions $\Gamma_v(\tau,\theta_i)$ (top) and $\Gamma_z(\tau,\theta)$ (middle) for different $\theta_i$. The bottom panel shows the number of points used in the statistical ensemble average of the correlation functions for given time-lag $\tau$. All plotted for different binned values of $\theta_i$.
    }
     \label{fig:fig2}
\end{figure}

Figure~\ref{fig:fig2} shows the curves of the normalized correlation functions $\Gamma_v(\tau,\theta_i)=C_{v}(\tau,\theta_i)/C_{v}(0,\theta_i)$ and  $\Gamma_z(\tau, \theta_i)=C_z(\tau, \theta_i)/C_z(0, \theta_i)$  as a function of time-lag $\tau$. For all $\theta_i$ the normalized correlations $\Gamma_v$ and $\Gamma_z$ drop sharply for $\tau\lesssim 12$ min, and they practically vanish when $\tau\gtrsim 30$ min. As it is shown in the bottom panel of Figure~\ref{fig:fig2}, all the correlations, except those corresponding to $\theta_4=60^\circ$,  were  measured with reasonably good statistics, with a minimum number of points higher than 100. Therefore, we do not consider the analysis for $\theta_4=60^\circ$ due to a lack of reliable statistics.

\begin{figure}
    \centering
    \includegraphics[width=0.5\textwidth]{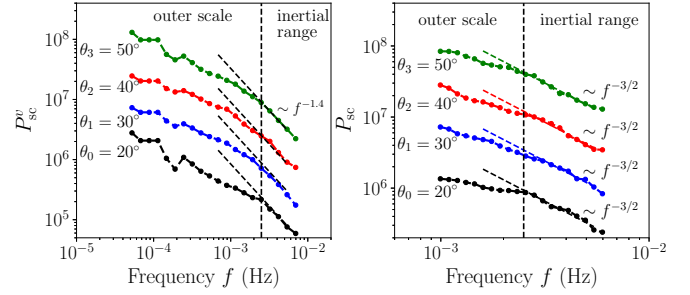}
       \vspace*{-0.5cm}
    \caption{Spacecraft-frame power spectra for velocity $P^{v}_{\sc}(\omega,\theta_i)$ (left panel) and for Elsasser field $\vec z^-$, $P_{\sc}(\omega,\theta_i)$ (right panel). All power spectra were plotted as a function the frequency $f=\omega/(2\pi)$ measured in the spacecraft frame. All spectra were fitted to power-law functions within the frequency range $[2.5\times 10^{-3},6\times 10^{-3}]$~Hz (inertial range). The different curves have been offset vertically for easier viewing. We also excluded the part of the power spectra that are affected by noise for $f> 6\times10^{-3}$ Hz. 
    }
    \label{fig:fig3}
\end{figure}

Figure~\ref{fig:fig3} displays power spectra $P^{v}_{\sc}(\omega,\theta_i)$ and $P_{\sc}(\omega,\theta_i)$ computed through the Fourier transform of the corresponding correlation functions (Figure~\ref{fig:fig2}) for each angle bin $\theta_i$. The four spectra in this figure have been artificially re-scaled to allow for better comparisons. We excluded the part of the power spectra (for $f\gtrsim 6\times 10^{-3}$) that is affected by the noise due to the time resolution of the plasma experiment. Also, we did plot the very low frequency part only for the power spectra $P^{v}_{\sc}(\omega,\theta_i)$ as we will use it to estimate the \rms~of the outer-scale fluid velocity.

The power spectrum $P^{v}_{\sc}(\omega,\theta_i)$ seems to steepen for frequencies above  $\sim 2.5\times 10^{-3}$ Hz showing a spectral index of about 1.4. The outer scale of the velocity field (frequency below $f_0\sim 2.5\times 10^{-3}$ Hz) follows a power law that is comparable to or steeper than $f^{-1}$. We estimate the value of the outer-scale~\rms~bulk speed, $\delta u_{i,0}$, for each power spectrum $P^{v}_{\sc}(\omega,\theta_i)$ as 
\begin{equation}
 \delta u_{i,0}=4\pi\int^{f_0}_0 P^{v}_{\sc}(2\pi f,\theta_i) df.
 \label{eq:du0}
\end{equation}
The right panel of Figure~\ref{fig:fig3} displays the power spectrum $P_{\sc}(\omega,\theta_i)$ within frequency range between $10^{-3}$ Hz and $10^{-2}$ Hz. The power spectra are fitted to power-laws in the inertial range (within the frequency range $f\in[2.5\times 10^{-3},6\times 10^{-3}]$~Hz), of the form $P_\sc(\omega,\theta_i)=D_i \omega^{-\alpha}$ for both $\alpha=3/2$ and $\alpha=5/3$. The values obtained for the energy constant $D_i$ for each value of $\alpha$ are listed in Table~\ref{tab:tab1}.


\begin{figure}
    \centering
    \includegraphics[width=0.425\textwidth]{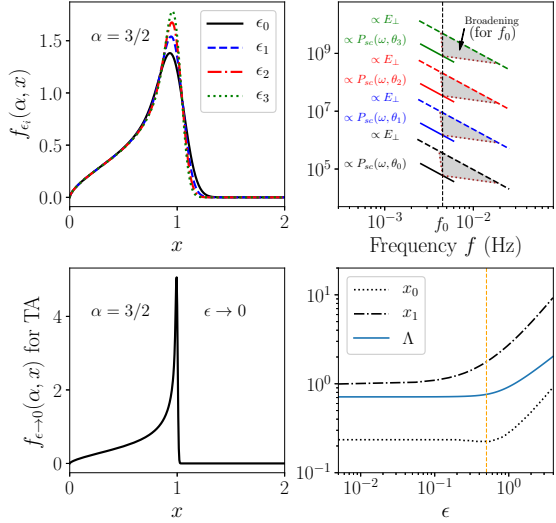}
    \vspace*{-0.1cm}
    \caption{ 
    Top left: The functions $f_{\epsilon_i}(\alpha_i,x)$ plotted as a function of the dimensionless variable $x$ for the empirical values of $\epsilon_i$ and $\alpha=3/2$. Top right: Part of the inertial-range power-law fits $\propto D_i (2\pi f)^{-\alpha_i}$ (colored solid lines) that correspond to $P_{i,\sc}$. The dashed colored lines are the corresponding reduced power spectra $\propto C_i f^{-\alpha_i}$ (where $f=k_\perp V_{i,\perp}/(2\pi)$). The different power-law curves have been offset vertically for easier viewing. The broadening in $k_\perp$ that corresponds to frequency $f_0=4.5\times 10^{-3}$ Hz is shown in the frequency-domain interval (gray areas) as $[k_{i,min},k_{i,max}] V_{i,\perp}/(2\pi)=[f_0/x_{i,1},f_0/x_{i,0}]\simeq[f_0/0.2, f_0/1.1]$ (see Eq.\ref{eq:broad}). Bottom left: Curves represent the function $f_{\epsilon\rightarrow 0}(\alpha,x)$ plotted for $\alpha=3/2$. Bottom right: Curves of $\Lambda$ (solid line), $x_0$ (dot line) and $x_1$ (dash line) plotted versus $\epsilon$ all estimated for $\alpha=3/2$.  }
    \label{fig:fig4}
\end{figure}
\section{The reduced spatial power spectrum}

\subsection{Derivation of $E(k_\perp)$ using BP19 model}

We estimate a reduced energy spectrum $E(k_\perp)$ associated with each measured spacecraft-frame $P_{\sc}(\omega,\theta_i)$. According to the BP19 model, for the strong turbulence case, the spectral index of the reduced power spectrum $E(k_\perp)$ in the inertial range will be the same as the spectral index of their corresponding frequency spectrum $P_{\sc}(\omega,\theta_i)$, therefore $E(k_\perp)=C_i k^{-\alpha}_\perp$ where $C_i$ is expected to be the same constant for each sampling angle $\theta_i$ if the turbulence is strong and highly anisotropic. From the BP19 model we have the following relationship
\begin{equation}
    P_{\sc}(\omega,\theta_i)=\Lambda_i\frac{C_i}{V_{i,\perp}}\left (\frac{\omega}{V_{i,\perp}}\right)^{-\alpha} 
    \label{Eq.Psc}
\end{equation}
where
\begin{eqnarray}
\Lambda_i&=&\int^{\infty}_0 f_{\epsilon_i}(\alpha,x) dx,\label{Eq.lambda} \\
   f_{\epsilon_i}(\alpha,x)&=&x^{\alpha-1} \bar{g}_{\epsilon_i}(x)
   \label{Eq.lambda2}
\end{eqnarray}
and $\bar{g}_\epsilon(x)$ is a function of the dimensionless parameter $x=\omega/k_\perp V_{i,\perp}$ that is connected to the probability distribution of function $g(y)$ (assumed to be a Gaussian distribution) as follows
\begin{equation}
  \bar{g}_{\epsilon_i}(x)=\frac{2}{\pi}\int^\pi_{0}\frac{1}{\epsilon_i} g\left(\frac{x+\cos\phi}{\epsilon_i}\right) d\phi.
  \label{Eq.gbar}
\end{equation}
where
\begin{equation}
 g(y)= \frac{1}{\sqrt{2\pi}} e^{-\frac{1}{2}y^2}
  \label{Eq.g}
\end{equation}
and the parameter $\epsilon_i=\delta u_{i,0}/(\sqrt 2 V_{i,\perp})$, with $V_{i,\perp}=\overline{V}_{i,0} \sin{\theta_i}$ is the field-perpendicular velocity of the spacecraft as seen in the plasma frame. All the empirical values of the above parameters are given in table 1.  The angle $\phi$ in Eq.~\eqref{Eq.gbar} is the direction of the wavevector ${\bf k}_\perp$ in the field-perpendicular plane. By replacing the power law-fits $P_{\sc}(\omega,\theta_i)=D_i \omega^{-\alpha}$ in Eq.~\eqref{Eq.Psc}, we get
\begin{equation}
C_i = \frac{D_i}{\Lambda_i} V^{-\alpha_i +1}_{i,\perp}
\label{Eq.C}
\end{equation}
Equation~\eqref{Eq.C} can now be used to find the values of $C_i$, summarized table~\ref{tab:tab1} that correspond to the reduced power spectra $E(k_\perp)$ in the inertial range. Interestingly, the values of $C_i$ are all around $6\times 10^{5}$ ($6\times 10^{4}$) in SI units for $\alpha=3/2$ ($\alpha=5/3$), and there is no dependency on the sampling angle $\theta_i$. This is a strong signature that the turbulence is strong and anisotropic as found in many previous works~\citep{horbury08,podesta09,chen11a}.

The upper left of Figure~\ref{fig:fig4} displays the function  $f_{\epsilon_i}(\alpha,x)$ for the empirical parameters $\epsilon_i$. The values of $\Lambda_i$ were obtained through Eq.~\eqref{Eq.lambda} and summarized in Table~\ref{tab:tab1}. The curves of $f_{\epsilon_i}(\alpha,x)$ are seen to be broad in $x=\omega/k_\perp V_{i,\perp}$. According to BP19, TA can be recovered when $\epsilon \ll 1$ and its accuracy worsens when the broadening of $f_{\epsilon_i}(\alpha,x)$ becomes significant,  as we discuss in the next subsection. The broadening in $x$ will basically lead to broadening in the field-perpendicular wavenumber $k_\perp$. 
This means that the energy $P_{\sc}(\omega,\theta_i)$ at a small frequency bin around $\omega$ corresponds to the energy in the wavenumber range $\Delta k_{i,\perp}=[k_{i,\min},k_{i,\max}]$ according to the energy power $E(k_\perp)$.  
This broadening can be determined from the broadening $\Delta x_i= [x_{i,0},x_{i,1}]$ of the function $f_{\epsilon_i}(\alpha,x)$. The values of $x_{i,0}$ and $x_{i,1}$ can be estimated from the following two prescriptions, 1) the integral 
 \begin{equation}
 \int^{{x_{i,1}}}_{x_{i,0}} f_{\epsilon_i}(\alpha,x) dx=\eta\Lambda_i
 \label{eq:int}
\end{equation}
 captures a desired fraction (e.g., $\eta\simeq 0.9$) of the $\Lambda_i$ parameter, and 2)
  \begin{equation}
 \int^{0}_{x_{i,0}} f_{\epsilon_i}(\alpha,x) dx= \int_{x_{i,1}}^{\infty} f_{\epsilon_i}(\alpha,x) dx.
 \label{eq:area}
\end{equation}
 With this prescription one then obtains the broadening in $k_\perp$ for each frequency as 
  \begin{eqnarray}
 k_{i,\min}=\omega/(x_{i,1}V_{i,\perp})\mbox{~and ~}k_{i,\max}=\omega/(x_{i,0}V_{i,\perp})
 \label{eq:broad}
  \end{eqnarray}
 corresponding to each power spectrum $P_{i,\sc}(\omega,\theta_i)$. Using $\eta=0.90$ in this analysis leads to the values of $x_{i,0}$ and $x_{i,1}$ summarized in Table~\ref{tab:tab1}. For all considered $\theta_i$, the broadening seems to be the same for the empirical values of $\epsilon_i\lesssim 0.1$.

In the upper right panel of figure \ref{fig:fig4} we illustrate\footnote{In the figure we re-scaled $E(k_\perp)$ by various factors for clarity} the broadening in $k_\perp$ that contributes to the power spectrum $P_{\sc}(\omega_0,\theta_i)$ for frequency $\omega_0=2\pi f_0$, where $f_0=4.5\times 10^{-3}$ Hz. It is worth mentioning that the methodology proposed in BP19 can be used to reconstruct the reduced energy spectrum as long as  $\tan{\theta_b}\gtrsim \delta u_0/v_A$, i.e.,  $\tan{\theta_b}\gtrsim 20^\circ$ for the data set considered in this work.

\begin{table}[!t]
\begin{center}
\begin{tabular}{c|cccc } 
 \hline
 $i$ & 0  & 1& 2 &3 \\
$\theta_i$ & 20$^\circ$  & 30$^\circ$ & 40$^\circ$ & 50$^\circ$ \\
 \hline
$V_{i,\perp}$ (km/s)&219&  319& 404& 482\\
$D_i\times 10^{-8}$ (SI) ($\alpha=5/3$)&1.5 &2.2 &2.3 & 2.7\\
$D_i\times 10^{-8}$ (SI) ($\alpha=3/2$)&2.6 &3.3 &3.8 & 4.6\\
$\delta u_{i,0}$ (km/s) & 32&38 &41 &43 \\
$\epsilon_i$ &0.10 &0.08 &0.07 &0.06   \\ 
$\Lambda_i$ ($\alpha=5/3$) &0.71 &0.71 &0.71 & 0.71  \\
$\Lambda_{\rm TA}$ ($\alpha=5/3$)&0.71 &0.71 &0.71 & 0.71  \\
$C_i \times 10^{-4}$ (SI) ($\alpha=5/3$)&6.0 &5.8 &5.7 &6.0 \\
$C_i \times 10^{-5}$ (SI) ($\alpha=3/2$)&6.0 &6.0 &6.2 &6.7 \\
$C_{i,\rm TA}\times 10^{-4}$ (SI) ($\alpha=5/3$)&6.0 &5.8 &5.7 &6.0 \\
$C_{i,\rm TA}\times 10^{-5}$ (SI) ($\alpha=3/2$)&6.0 &6.0 &6.2 &6.7 \\
$x_{i,0}$ ($\alpha=5/3,3/2$) & 0.2& 0.2 &0.2 & 0.2 \\
$x_{i,1}$ ($\alpha=5/3,3/2$) & 1.1 &  1.1& 1.0& 1.0\\

 \hline
\end{tabular}
\caption{Relevant measured parameters used to reconstruct power spectrum $E(k_\perp)$ from BP19 model. The mean Alfv\'en velocity~$v_A=80$~km/s.}\label{tab:tab1}
\end{center}
\end{table}

\subsection{Derivation of $E(k_\perp)$ using the Taylor approximation (TA)}
As suggested in BP19 sweeping model, the Taylor approximation can be recovered in the limit when ${\epsilon\xrightarrow{}0}$. It is straightforward to show from Eq.~\eqref{Eq.gbar}
\begin{equation}
    \lim_{\epsilon\rightarrow 0}\bar g_\epsilon(x)=\left\{\begin{array}{ccc}
    \frac 2{\pi\sqrt{1-x^2}} & 0<x<1\\
    0                     &\hbox{otherwise}
    \end{array}
    \right.
     \label{eq:gbar0}
\end{equation}
from where it follows that using equations~\eqref{Eq.lambda}~and~\eqref{Eq.lambda2}
\begin{equation}
    \Lambda_{\rm TA}=\frac 2\pi\int_0^1\frac{x^{\alpha-1}}{\sqrt{1-x^2}}dx=\frac{\Gamma\left(\frac{\alpha}2\right)}{\sqrt\pi\Gamma\left(\frac{\alpha+1}2\right)}
\end{equation}
where $\Gamma(x)$ is the Gamma function. Thus the relationship that connects the reduced power spectrum $E(k_\perp)\simeq C_{i,\rm TA} k_\perp^{-\alpha_i}$ and the spacecraft-frame power spectrum $P_{\sc}(\omega,\theta_i)$ given in Eq.~\eqref{Eq.Psc} will become

\begin{equation}
    P_{\sc}(\omega)=\Lambda_{\rm TA} \frac{C_{i,\rm TA}}{V_{i,\perp}}\left(\frac{\omega}{V_{i,\perp}}\right)^{-\alpha}
    \label{Eq.Pscth}
\end{equation}

The values of $\Lambda_{\rm TA}$ and the corresponding energy constants $C_{i,\rm TA}$ for the spectral indices $\alpha=3/2$ and $5/3$ are listed in Table~\ref{tab:tab1}. The results from this analysis show that for  $C_i\simeq C_{i,\rm TA}$ in SI units (and  $\Lambda_i \simeq \Lambda_{\rm TA}$) for both values of $\alpha$, suggesting that the TA is still a good approximation for empirical values of $\epsilon_i\lesssim 0.1$. The function $f_{\epsilon\rightarrow 0}(\alpha=3/2,x)$ that is used to compute $\bar g_{\epsilon\rightarrow 0}(x)$ in Eq.~\eqref{eq:gbar0} in TA are plotted in the lower left panel of figure \ref{fig:fig4}. Even when $\epsilon\ll 1$ (TA) there is still some broadening that is caused by the integration over angle $\phi$. The estimation of the broadening in $x$ for the TA case provide the same values of $x_{i,0}$ and $x_{i,1}$ as found using BP19 methodology.

The analysis we present suggests that the BP19 model and TA provide a similar prediction for the energy constant. However, the TA obtained from BP19 in the limit of $\epsilon\rightarrow 0$ takes into account the broadening in $k_\perp$ for a corresponding angular frequency $\omega$, which arises from the angular integration of the wavevector in the field-perpendicular plane.  To the best of our knowledge, the effect of this broadening within TA approximation has not been taken into account in solar wind observations. 
As our results show, for observations with $\epsilon\lesssim 0.1$, both the energy constant and broadening are the same as with the TA. For larger values of $\epsilon$, this is not necessarily the case.  For instance, in the lower right panel of Figure \ref{fig:fig4} we estimate the parameter $\Lambda$ and the corresponding values of $x_0$ and $x_1$ assuming $\alpha=3/2$ and varying $\epsilon$ from $10^{-3}$ to 4. Interestingly, the parameter $\Lambda$ remains roughly constant for $\epsilon\lesssim 0.5$, which means that we would expect the same values for the energy constant whether applying BP19 model or TA for this range of $\epsilon$. The parameter $\Lambda$ begins to change appreciably when $\epsilon\gtrsim 0.5$. This value of $\epsilon\gtrsim0.5$ might be obtained near the sun region when dealing with time signals near the sun region where \emph{PSP} is going to explore. The broadening seems to not change dramatically when $\epsilon\lesssim 0.1$, however, when $\epsilon > 0.1$ the curves of $x_0$ and $x_1$ begin to change. We conclude that for $\epsilon\lesssim0.1$ both TA and BP19 lead to the same energy constant and broadening, for $0.1\lesssim\epsilon\lesssim0.5$ the TA properly captures the energy constant but not the broadening, while for any value of $\epsilon> 0.5$ the TA can no longer be justified.


\section{Conclusion} \label{sec:conclusion}

In this analysis we applied the methodology proposed recently by Bourouaine and Perez (2019) (BP19) to reproduce the reduced power spectra  $E(k_\perp)$ in the inertial range  (in the plasma frame of reference) from the empirically measured power spectra $P_{\sc}(\omega,\theta_i)$ for each binned angle $\theta_i$. The values of the constant $C_i$ seems to be unchanged with respect to the sampling angle $\theta_i$. This conclusion is clearly consistent with the fact that the studied turbulence is strongly anisotropic. Interestingly, we found that, when $\epsilon\lesssim 0.1$, the estimated energy constant $C_i$ from BP19 model are comparable to the one obtained through TA, but at any value of $\epsilon$, including when $\epsilon\rightarrow 0$ (for TA), there will be always a significant broadening in $k_\perp $ associated with a given frequency $\omega$. The broadening in $k_\perp$ that appears when $\epsilon\rightarrow 0$ is due to the integration over $\phi$ angle. Many previous works considered the integration over $\phi$ using TA in the estimation of the energy spectrum $E(k_\perp)$~\citep[e.g.,][]{bourouaine13,vech17,martinovic19}.  Broadening due to sweeping of large-scale will be more significant as $\epsilon$ increases. The application of BP19 model provides a significant difference in the evaluation of the energy constant $C$ than when TA is used if $\epsilon$ is larger than 0.5, which may very well occur in the solar wind namely near the sun region where \emph{PSP} is expected to explore.

\acknowledgments

\acknowledgments
SB was supported by NASA grants NNX16AH92G, 80NSSC19K0275 and 80NSSC19K1390. JCP was partially supported by NASA grants NNX16AH92G, 80NSSC19K0275 and NSF grant AGS-1752827. The authors also like to acknowledge constructive discussions with Kristopher Klein, Mihailo Martinovi\'c and Christopher Chen. 

\bibliography{MyLibrary.bib}






\end{document}